\documentclass[journal]{IEEEtran}
\IEEEoverridecommandlockouts
\usepackage{cite}
\usepackage{amsmath,amssymb,amsfonts}

\usepackage{graphicx}
\usepackage{subfigure}
\usepackage{textcomp}
\usepackage{xcolor}
\usepackage{multirow}
\usepackage{array}
\usepackage{hyperref}
\usepackage{algorithm}
\usepackage{algpseudocode}
\usepackage{graphicx}

\usepackage{makecell} 
\usepackage{booktabs} 
\usepackage{multirow}   

\def\BibTeX{{\rm B\kern-.05em{\sc i\kern-.025em b}\kern-.08em
    T\kern-.1667em\lower.7ex\hbox{E}\kern-.125emX}}
\begin{document}

\title{
Four-Transistor Bipolar Series-Parallel Module Structure for Cascaded Bridge and Modular Multilevel Circuits
}

\author{\IEEEauthorblockN{Jinshui Zhang}
and
\IEEEauthorblockN{Stefan M Goetz}
}

\maketitle

\begin{abstract}
    With their great scalability and flexibility, cascaded-bridge and modular multilevel converters have enabled a variety of energy applications, such as offshore wind power, high-voltage dc power transmission, power-quality management, and cutting-edge medical instrumentation. The incorporation of parallel connectivity between modules equips systems with advantages such as sensorless balancing, switched-capacitor energy exchange, and reduced impedance. However, existing topologies require many individual switches---eight transistors per module. Efforts to use fewer switches, instead, have previously compromised their functionality. We propose a new module topology, named the direction-selective parallel (DiSeP) structure, which requires only four transistors per module---the same as an H bridge---but can achieve bidirectional equilibration, bipolar module output, and inter-module switched-capacitor features.
    
    This topology is highly attractive for existing converters with cascaded bridge elements, as the addition of only four diodes enables key features such as sensorless balancing and inter-module energy exchange. Thus, the module can outcompete H bridges in their applications, as it adds parallel modes without any additional transistors. Compared to double-H bridges (CH²B), it saves as many as half of the transistors.
    We elaborate on its working principles and key design considerations. We validate our theories on an experimental prototype with six modules. This prototype attains a total voltage harmonic distortion plus noise (THD+N) of 10.3\% and a peak efficiency of 96.3\%. Furthermore, the modules achieve autonomous sensorless balancing under open-loop control.
\end{abstract}



\section{Introduction}
Multicell converters, such as cascaded bridge converters (CBC) and modular multilevel converters (MMC), have enabled various unprecedented systems, such as offshore and onshore wind converters, high-voltage direct current (HVDC) systems, reactive power as well as harmonics compensation, and medium-voltage motor drives \cite{4591920, 5544594, 5279054, 8601394, 6342749, 6231806}. Their capability of integrating high quality, high power, and wide bandwidth has furthermore established them outside conventional energy systems, such as medical devices, pulse generators, and neuroscientific instrumentation \cite{goetz2017development, li_av, 6347016, lz_j_p_medi, 9097830}.

CBC\,/\,MMC modules typically consist of one energy storage element, such as batteries or capacitors, as well as a set of transistors to loop the energy storage element into the output or out of it. These modules were originally limited to series and bypass connectivity \cite{lesnicar2003innovative, marquardt2010modular, glinka2005new}. Modules in series mode increase the output voltage by one level and direct the output current through their capacitors. Those in bypass mode guide the current around their capacitors to the next module.

The switching of modules between active series, in which the module capacitors charge or discharge, and bypass, leads to several challenges. The capacitors need to be large enough so that their charge can provide or absorb the load current sufficiently long, typically on the order of a period of the sinusoidal output. However, the capacitor voltages of the modules tend to drift apart if their individual net charge flow is not perfectly equal over time. This deviation, affected by the current direction and amplitude as well as the module capacitance, can lead to output distortion due to uneven output steps. Severe situations may put modules into the undervoltage or overvoltage regions, which endangers the health of the hardware. The converter controller has to actively balance the capacitors to reduce the voltage spread. In addition, there is a strong desire to reduce the module capacitance as the capacitors take up a large volume so that balancing should keep the capacitor voltage even closer together despite reduced capacitance \cite{7362232,10.1049/iet-pel.2014.0328}.

Furthermore, a bypassed module contributes nothing but impedance and loss to the system. This inefficiency becomes particularly significant when the modulation index is low, e.g., in variable-speed motor drives.

A large body of literature attempts to solve this problem with control schemes \cite{9152155, 9815810, 9328618, 8626987,8013775,8341622}. Most of these balancing control schemes require lots of sensors as estimators turned out to be vulnerable and cannot provide software-free, hardware-safe operation. The introduction of switched-capacitor features for parallel inter-module connectivity in addition to the series mode can solve most of such issues.

Instead of bypassing modules as in conventional cascaded bridges, the converter can parallel the capacitors of two or more neighbors to transfer energy or to just equalize capacitors. This dynamically joint module group collects the total capacitance of all paralleled modules and can temporarily operate as one module with larger capacitance, lower source impedance, and reduced voltage ripple until they are dynamically separated again \cite{6763109, 9415177, 7468193, 9109731, 8571242}. 
More importantly, the parallel connection between modules ensures ideal voltage sharing without extra control or sampling needs \cite{8571242,7801087}. Module balancing is now guaranteed by hardware, which greatly improves system reliability and resilience \cite{zhang2024frequency}.

There are different ways to implement parallel connectivity in MMC\,/\,CBC converters \cite{9415177}. Excluding the discussion on local parallelization of capacitors within the same module \cite{Bharath_2022,7167913}, which does not offer the aforementioned advantages, key topologies allowing inter-module parallelization can be narrowed down to cascaded double-H-bridge (CH2B) \cite{Goetz_2015}, symmetrical \cite{9344675} and asymmetrical \cite{Li_2019} double-half-bridge, as well as diode- and switch-clamped modules \cite{tashakor2020modular, yin2018modular, liu2016novel, zheng2017medium,9219153}. 

The CH2B topology as one of the earliest representatives of this family requires eight transistors in each module. The cost of transistors and accessories such as gate drivers is a leading challenge of this technology. Other topologies try to use fewer switches but sacrifice their functionalities. The asymmetrical structure, for instance, can only support unipolar output, whereas the symmetrical topology doubles the number of individual capacitors.

Diode- and switch-clamped circuits  reduce the number of transistors in series\,/\,parallel circuits, but lose module states.
Diode-clamped and various switch-clamped modules, for instance, offer a unidirectional parallel path \cite{tashakor2020modular, yin2018modular, liu2016novel, zheng2017medium}, which can simplify modular multilevel and cascaded bridge converter operation with minimum monitoring and balancing effort. As diode-clamped cascaded converters can only equalize modules unidirectionally from one end to the other, appropriate control can assign active power to the modules dependent on the position and form the equivalent of a \textit{bucket chain} \cite{9969159}. However, they fail to ensure theoretical ideal balancing without control assistance or sensors. 

In this paper, we propose a new module topology, a direction-selective parallel (DiSeP) or quad-chopper structure. This configuration uses only four switches per module---not even one transistor more than an H bridge---but in addition to all H-bridge module states it achieves bipolar output and bidirectional parallelization.

\section{Proposal of Direction-Selective Parallel Modules}
This section illustrates the structure of the DiSeP topology and its key features, such as impedance and the module balancing mechanism. We also suggest potential variations of this circuit and compatible modulation schemes.

\subsection{Circuit Structure and Working Modes}
\begin{figure}
    \centering
    \includegraphics{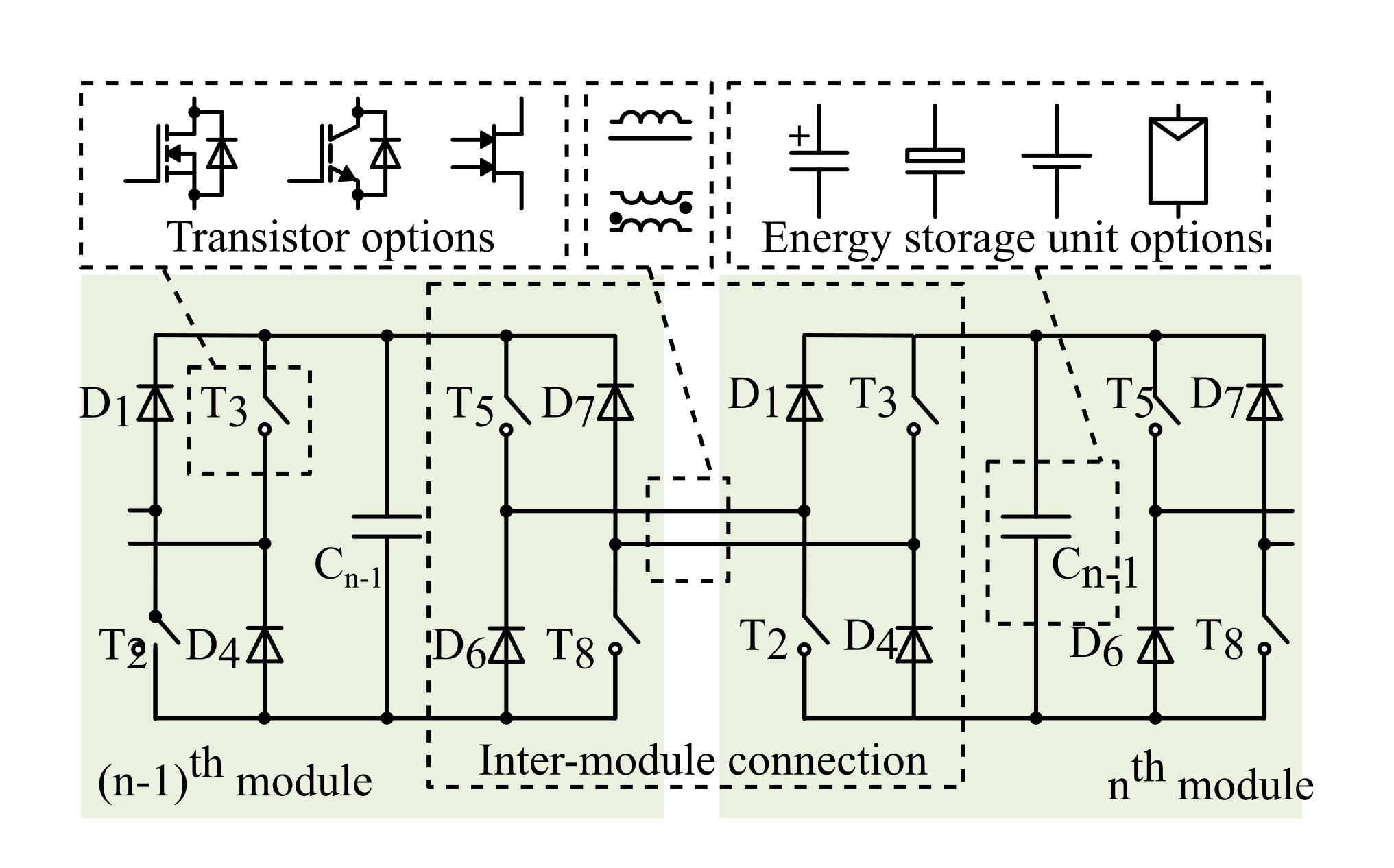}
    \caption{Circuit structure, connection pattern, and implementation of the DiSeP topology. Each module consists of four switches, four diodes, and one energy storage or source element. Switches can be implemented with any kind of transistor. The energy storage or source elements support various options such as polarized or unpolarized capacitors, batteries, or solar panels. Two adjacent modules are wired through two leads, which can introduce magnetics in between to adjust the differential- and\,/\,or the common-mode inductance.}
    \label{fig:disep_structure}
\end{figure}
\begin{figure*}[t]
    \centering
    \includegraphics[width=.9\textwidth]{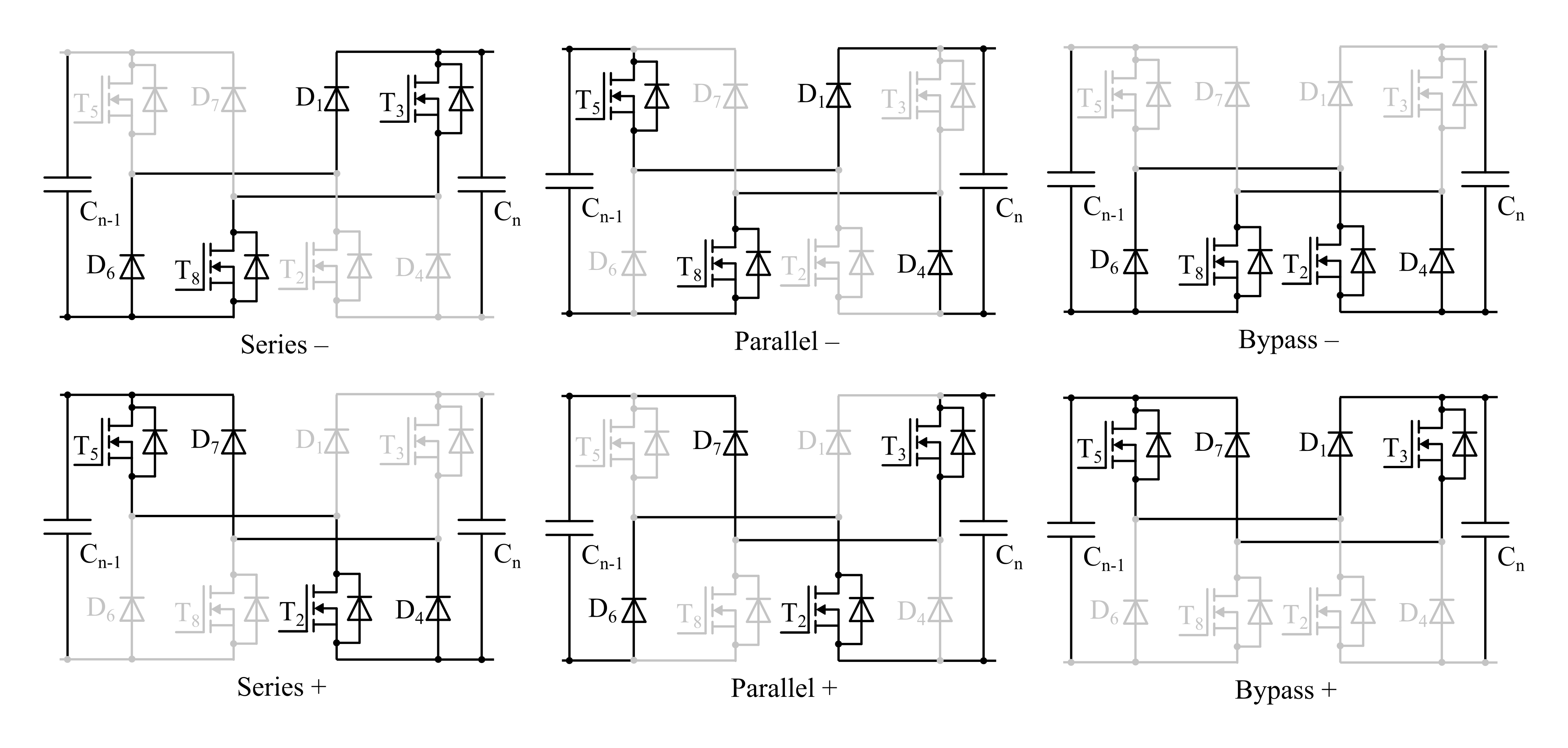}
    \caption{Module-interconnection states  of the DiSeP module.
             Series modes: \emph{Series--} connects the negative pole of $C_{n-1}$ to the positive pole of $C_n$, whereas \emph{Series+} connects the positive of $C_{n-1}$ to the negative pole of $C_n$. 
             Parallel modes: \emph{Parallel--} forms a clockwise parallelization loop to charge $C_n$, whereas \emph{Parallel+} forms a counterclockwise loop to charge $C_{n-1}$. 
             Bypass modes: \emph{Bypass--} allows a path through the negative rail of the modules, while \emph{Bypass+} does through the positive.}
    \label{fig:disep_working_modes}
\end{figure*}
Figure \ref{fig:disep_structure} illustrates the structure and connection pattern of DiSeP modules. Each module comprises one energy storage element and two types of half-bridges: diode--switch (D1--T2 and D7--T8) and switch--diode (T3--D4 and T5--D6) totaling four transistors and four diodes. Two pairs of diode--switch half-bridges introduce the first and fourth outreaching connections, while two pairs of switch--diode half-bridge module terminals account for the second and third. To connect two modules, the third half bridge of one module is wired to the first of the next, while the fourth to the second of the next. 
Thus, with no more transistors but only additional diodes, the module can do everything that H bridges can yet offers the additional features of parallel connectivity, such as switched-capacitor operation and sensorless balancing.

The energy storage element offers a variety of options, including polarized capacitors such as electrolytic capacitors, unpolarized capacitors such as film or ceramic capacitor, and batteries. We use a general capacitor model for demonstration. The switches can be implemented with current-bidirectional two-quadrant switches such as field-effect transistors (FETs), insulated-gate bipolar transistors (IGBTs), and high-electron-mobility transistors (HEMTs). In this paper, we demonstrate the circuit with FETs.

The DiSeP circuit supports six dynamic active module-interconnection modes (Figure \ref{fig:disep_working_modes}). The \emph{Series} modes connect the opposite poles of adjacent modules via two paralleled paths of two transistors and two diodes. The \emph{Parallel} modes form loops between neighboring modules for energy balancing. The two \emph{Parallel} modes support different directions of current flow and enable complementary bidirectional energy exchange. The \emph{Bypass} modes conduct the load current through either the positive or negative bus of the modules and contribute zero voltage to the output loop.

\subsection{Conducting Path and Impedance}
\begin{figure}[htbp]
    \centering
    \subfigure[\textit{Series--  }]{\includegraphics[width=0.4\textwidth]{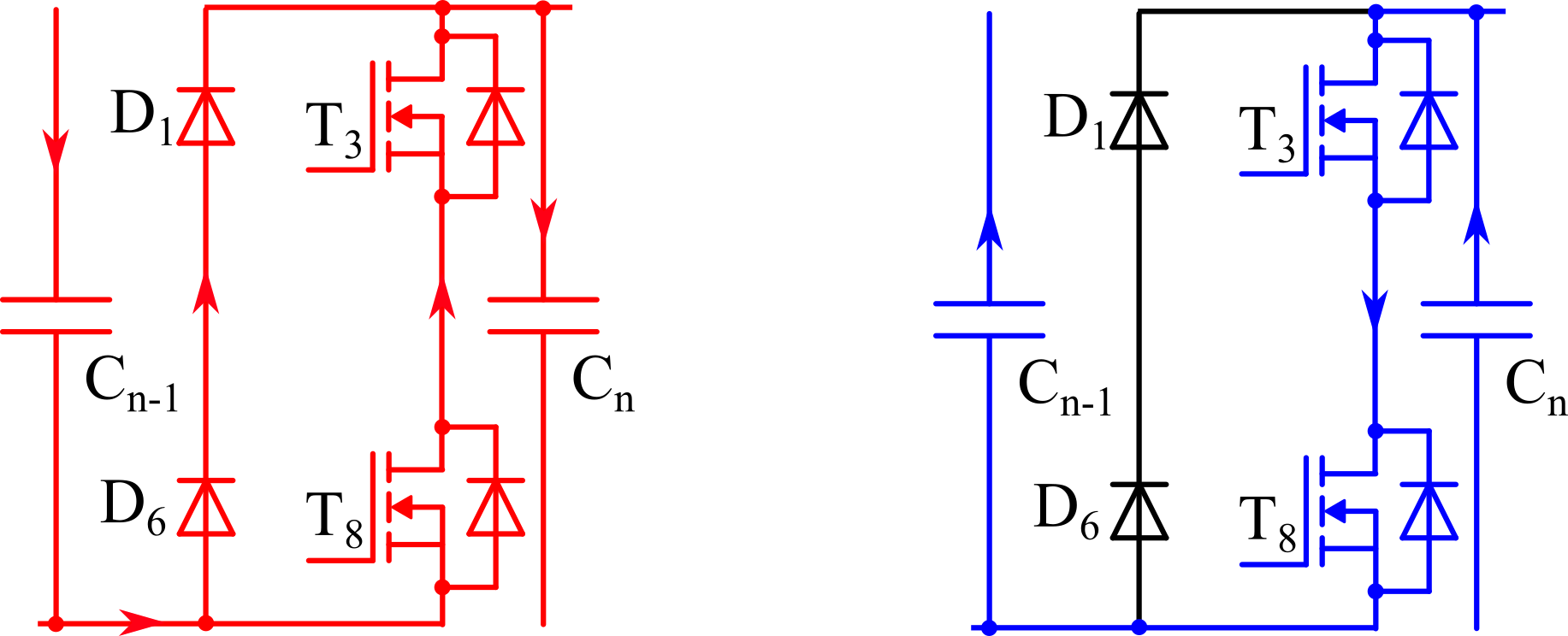}}
    \subfigure[\textit{Series+   }]{\includegraphics[width=0.4\textwidth]{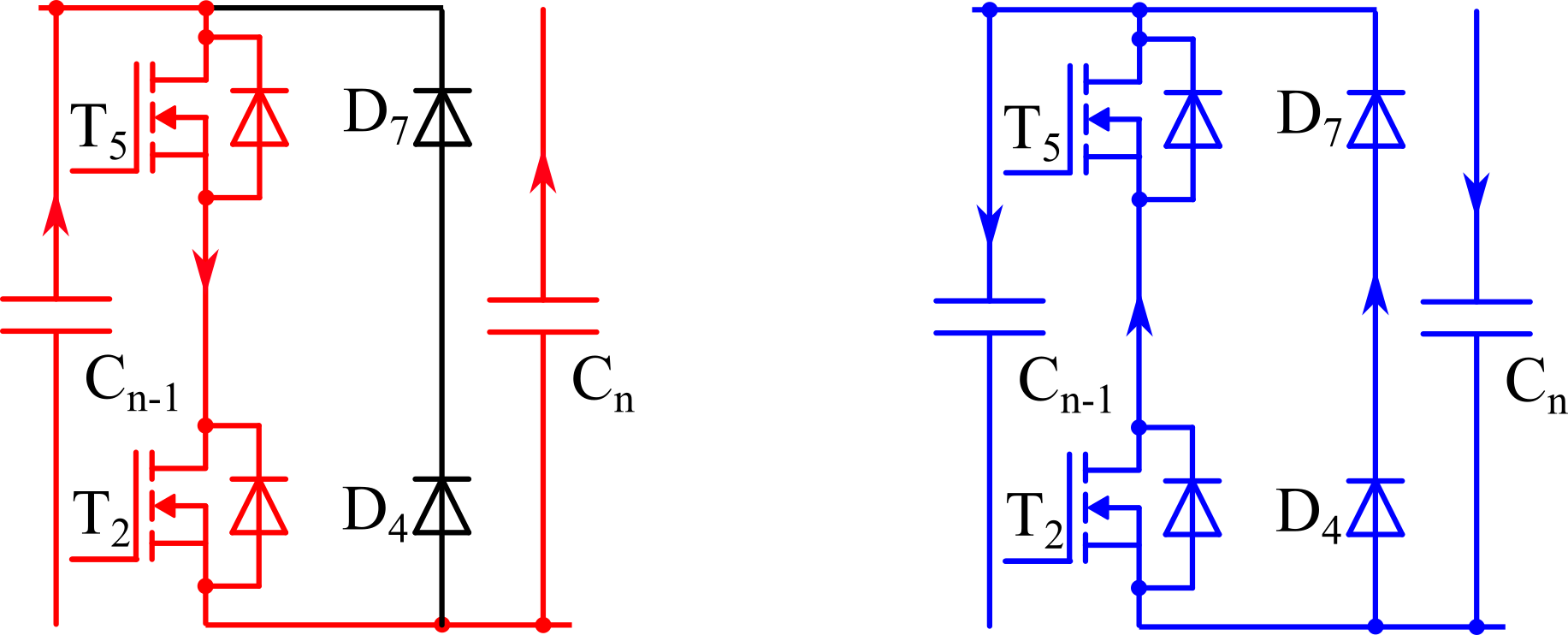}}
    \subfigure[\textit{Parallel--}]{\includegraphics[width=0.4\textwidth]{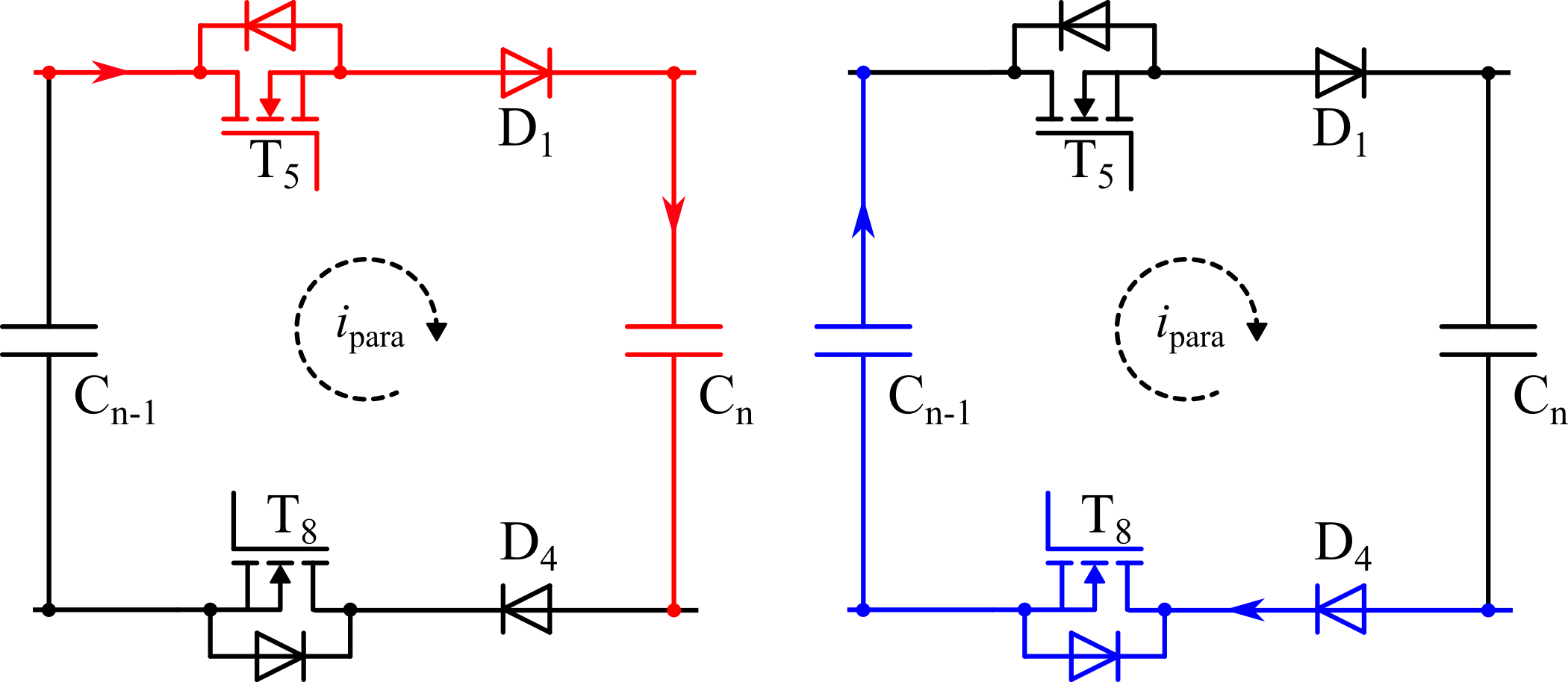}}
    \subfigure[\textit{Parallel+ }]{\includegraphics[width=0.4\textwidth]{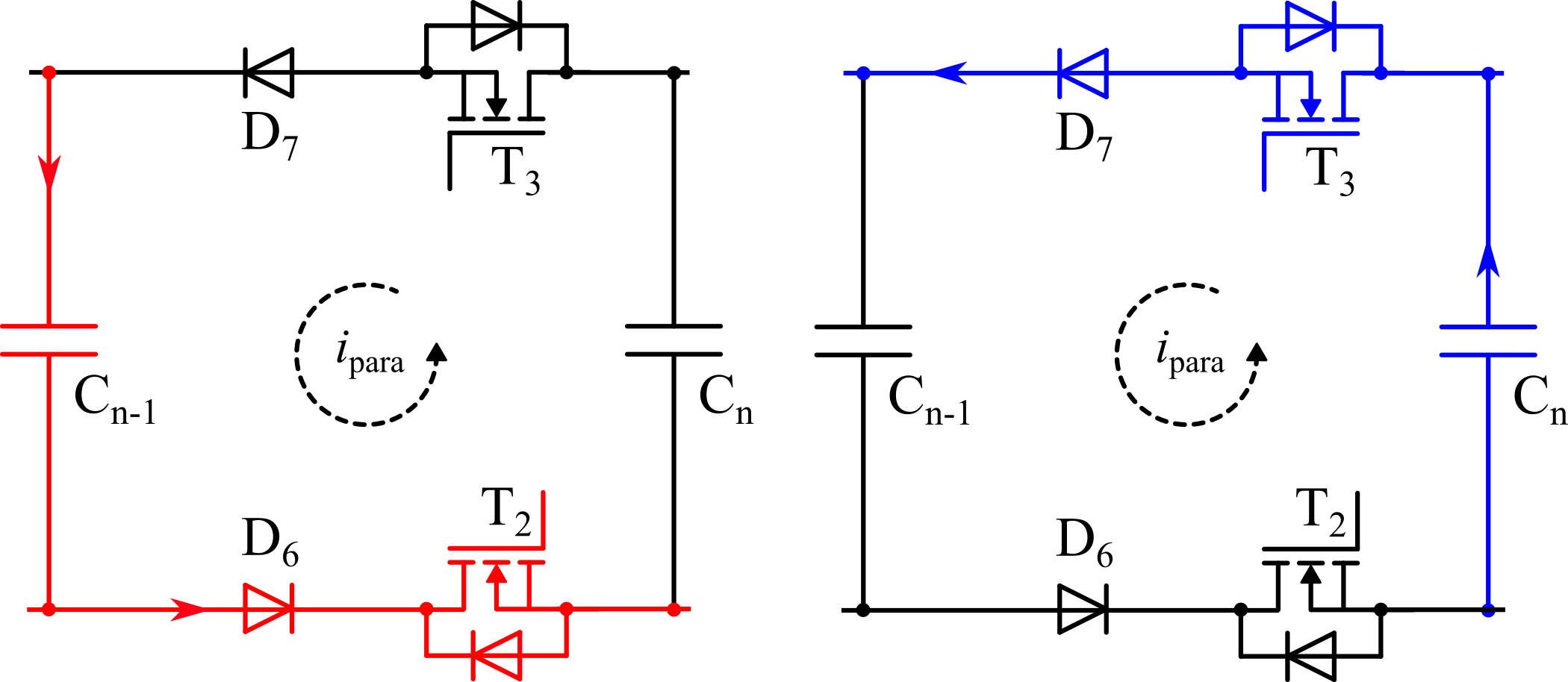}}
    \subfigure[\textit{Bypass--  }]{\includegraphics[width=0.4\textwidth]{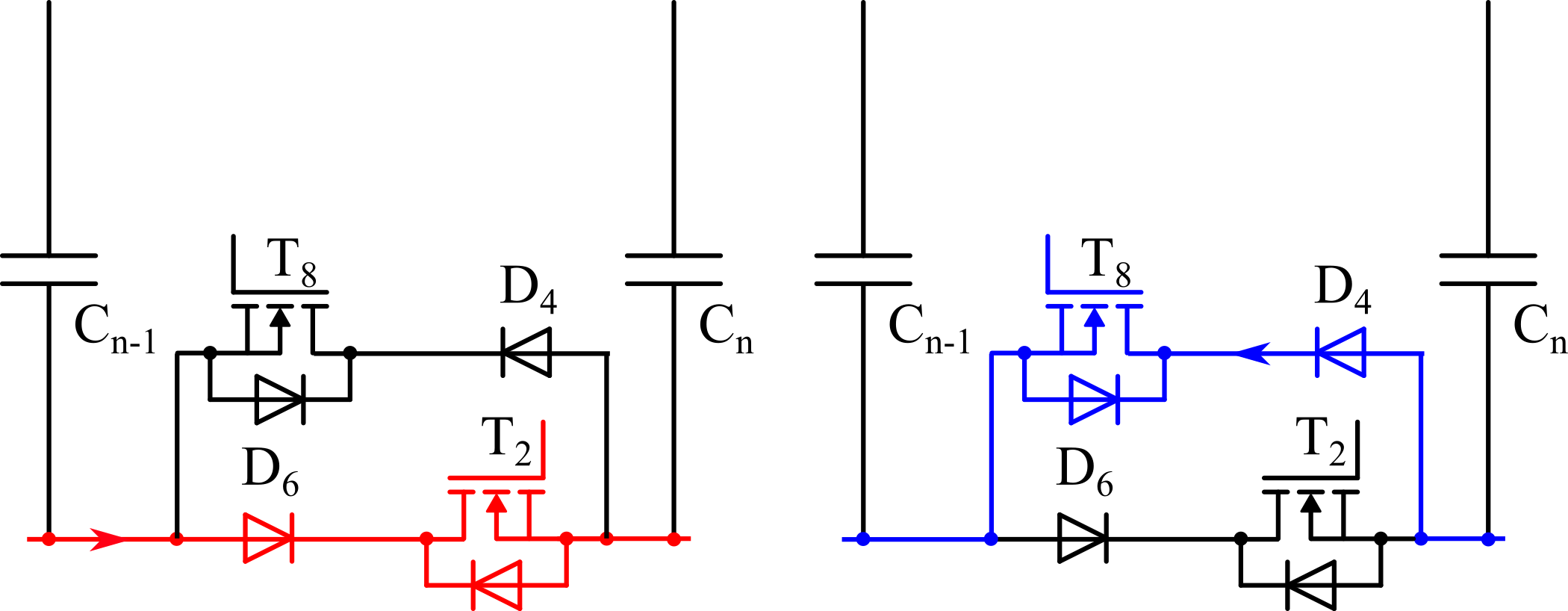}}
    \subfigure[\textit{Bypass+   }]{\includegraphics[width=0.4\textwidth]{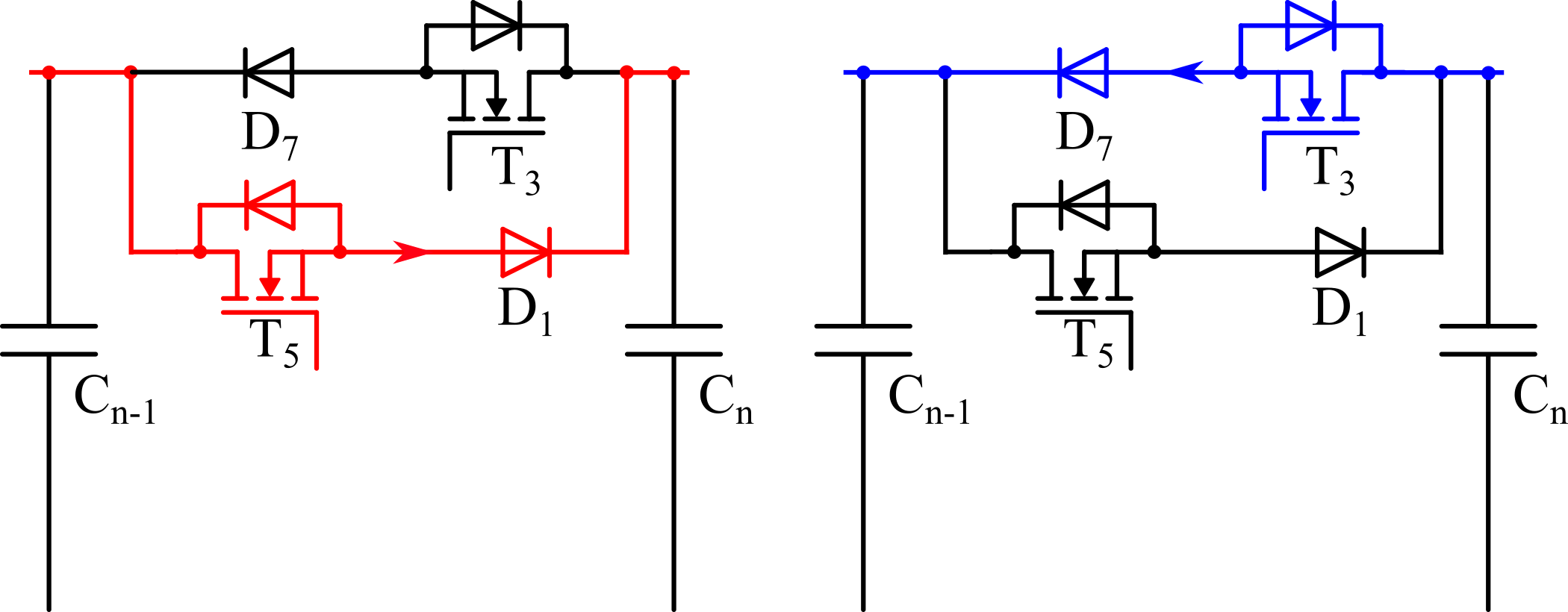}}
    \caption{Conduction paths of DiSeP inter-module connection in different states; the left-side demonstrates a current flowing from the $(n-1)^\textrm{th}$ module to the $n^\textrm{th}$, whereas the right-side shows the opposite.}
    \label{fig:current_path}
\end{figure}

Figure \ref{fig:current_path} illustrates the current paths of each mode. Based on the current paths, we can derive the impedance of each mode. In the \emph{Bypass} and \emph{Parallel} modes, DiSeP modules exhibit an equivalent impedance to the conventional H-bridge circuit---specifically a transistor and a diode in series. This impedance consists of one resistance and one PN junction for unipolar switches, such as FETs, and two PN junctions for bipolar devices, such as IGBTs. Improvements occur in the \emph{Series} modes. With a specific current direction, namingly current flowing from the $(n-1)^\textrm{th}$ module to the $n^\textrm{th}$ in \emph{Series--} mode and the opposite in \emph{Series+}, the current path comprises two paralleled branches, which consist of respectively two transistors and two diodes in series. 
The impedance of these paralleled branches follows 
\begin{equation}
    z_p = \frac{v_p}{i},
\end{equation}
where $v_p$ is the voltage of the paralleled branches and $i$ the current. These electric quantities follow Kirchhoff's circuit laws per
\begin{equation}
    \left\{
        \begin{aligned}
            & i_t + i_d = i,\\
            & v_t = V_\textrm{d} = v,
        \end{aligned}
        \right.
\end{equation}
where $i_d$ and $i_t$ are respectively the current flowing in the diode and transistor branches.
We adopted an ideal model for diode and FET according to 
\begin{equation}
    \begin{aligned}
        & v_\textrm{d} = V_\textrm{d}, \\
        & v_\textrm{t} = R_\textrm{ds,on} i_\textrm{t}.
    \end{aligned}
\end{equation}
The impedance obeys
\begin{equation}
    \begin{aligned}
    z_p & = \left\{
        \begin{aligned}
            & R_\textrm{ds,on}, & i < \frac{V_\textrm{d}}{R_\textrm{ds,on}} \\
            & \frac{V_\textrm{d}}{i}, & i \geq \frac{V_\textrm{d}}{R_\textrm{ds,on}}
        \end{aligned}
        \right. \\
        & = \text{min}\left(R_\textrm{ds,on}, \frac{V_\textrm{d}}{i}\right)\!,
    \end{aligned}
\end{equation}
which is the lower envelope of impedances of these two branches. However, most semiconductors exhibit a nonlinear \emph{i--v} relationship, a nonlinear simulation or direct measurement would be more accurate.

\subsection{Parallel Mechanism}
Although both DiSeP and CH2B can  bidirectionally parallelize modules, DiSeP uses a different mechanism. We illustrate these unique features from aspects of the direction selectivity, dynamics and energy loss. 

\subsubsection{Direction Selectivity}
Both of the parallel modes of the CH2B circuit can initiate a bidirectional equilibrium. By contrast, the DiSeP structure allows different directions with different parallel modes (Figure \ref{fig:current_path}). In the \emph{Parallel--} mode, capacitor $C_{n-1}$ charges $C_n$, whereas in the \emph{Parallel+} mode, the opposite occurs. These two parallel modes facilitate complementary parallelization directions. This mechanism underlies the name \emph{direction-selective parallel}.

\subsubsection{Dynamics of Parallelization}
\begin{figure}
    \centering
    \includegraphics{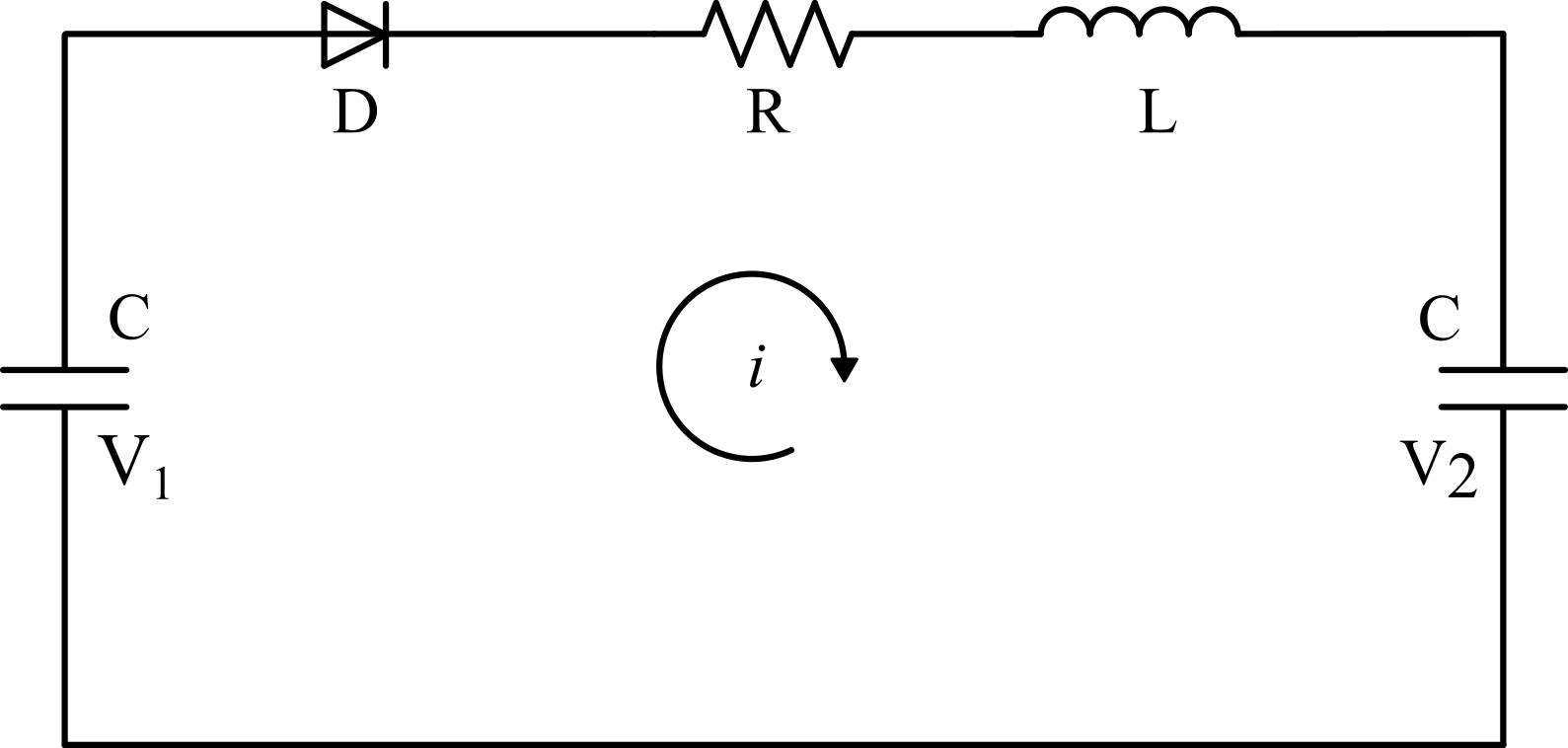}
    \caption{Equivalent circuit of a parallelization loop between two adjacent modules, which consists of two capacitors with the same capacitance but different initial voltage, one resistance that represents the parasitics and transistors, one diode that collectively represents the voltage drop of all diodes and transistors, one inductance from stray flux as well as optionally added magnetics.}
    \label{fig:paralleling_loop_stray_circuit}
\end{figure}
\begin{table*}[htbp]
    \centering
    \caption{Comparison of Paralleling Behavior under Different Circuit Configurations}
    \label{tab:comparison_paralleling_results}
    \begin{tabular}{c c c c}
        \toprule
        Circuit & Inter-Module Configuration & Steady-State Voltage Difference $\Delta V_{\infty}$ & Energy Loss $\Delta E$ \\
        \midrule
        CH2B & -- & 0 & $\frac{1}{4}C(V_1 - V_2)^2$ \\
        \midrule
        \multirow[b]{2}{*}{DiSeP} 
        & $R^2 \geq \frac{8L}{C}$ 
        & $V_\textrm{d}$ 
        & $\frac{1}{4}C\left((V_1 - V_2)^2 - V_\textrm{d}^2 \right)$ \\
        \cmidrule(l){2-4} 
        & $R^2 < \frac{8L}{C}$ 
        & \makecell{$-(V_1 - V_2)e^{\frac{\alpha \pi}{\beta}} + V_\textrm{d}\left(1 + e^{\frac{\alpha \pi}{\beta}}\right)$}
        & \makecell{
            $\frac{1}{4} C (V_1 - V_2 - V_\textrm{d})^2 \left(1 - e^{\frac{2\alpha \pi}{\beta}} \right) \times \left(1 + \frac{2V_\textrm{d}}{(V_1 - V_2 - V_\textrm{d})\left(1 - e^{\frac{\alpha \pi}{\beta}} \right)} \right)$
        } \\
        \bottomrule
    \end{tabular}
\end{table*}
We model the paralleling process with the circuit presented in Figure \ref{fig:paralleling_loop_stray_circuit}, where two capacitors have the same capacitance but different initial voltages $V_1, V_2$.  A condition of $V_1 > V_2 + V_\textrm{d}$ is necessary to initiate parallelization. Otherwise, the parallel mode automatically degrades to bypass modes. The diode entails a forward voltage drop of $V_\textrm{d}$. The inductance and resistance reflect parasitics, which can also include engineered magnetics \cite{zhang2024analytical}. For FET switches, their on-resistance can be incorporated into the effective resistance $R$. For IGBT switches, their on voltage drop can be merged into the diode behavior. Given such a circuit, the parallelization process can be described as
\begin{equation}
    \left\{
        \begin{aligned}
            & R i + L \frac{\textrm{d}i}{\textrm{d}t} + v_2 - v_1 + V_\textrm{d} = 0,\\
            & i = C \frac{\textrm{d}v_2}{\textrm{d}t} = -C \frac{\textrm{d}v_1}{\textrm{d}t},
        \end{aligned}
        \right.
\end{equation}
with an initial condition of 
\begin{equation}
    \left\{
\begin{aligned}
    v_1(0) & = V_1, \\
    v_2(0) & = V_2, \\
    i(0)   & = 0.
\end{aligned}
\right.
\end{equation}
We obtain the current model as 
\begin{equation}
    L \frac{\textrm{d}^2 i}{\textrm{d}t^2} + R \frac{\textrm{d}i}{\textrm{d}t} + \frac{2}{C} i = 0.
\end{equation}
Starting from an initial condition of zero current, the paralleling process ends at the next zero crossing of current due to the existence of the diode. We can obtain two solutions under different parameter configurations. \\
When the parameters fulfill $R^2 \geq \frac{8L}{C}$, i.e., the inter-module connection is resistance-dominated, the paralleling dynamics have a solution of 
\begin{equation}
    i(t) = \frac{V_1 - V_2 - V_\textrm{d}}{L \sqrt{\frac{R^2}{L^2} - \frac{8}{LC}}} \left(e^{r_1 t} - e^{r_2 t}\right),
\end{equation}
where 
\begin{equation}
\begin{aligned}
    & r_1 = -\frac{R}{2L} + \sqrt{\frac{R^2}{4L^2} - \frac{2}{LC}}, \\
    & r_2 = -\frac{R}{2L} - \sqrt{\frac{R^2}{4L^2} - \frac{2}{LC}}.
\end{aligned}
\end{equation}
Since $e^{r_1 t} - e^{r_2 t} \geq 0$, this scenario ends with a unipolar decaying of current. 
We obtain the post-parallelization voltage as
\begin{equation}
\begin{aligned}
    & v_{11}(\infty) = \frac{V_1 + V_2 + V_\textrm{d}}{2}, \\
    & v_{12}(\infty) = \frac{V_1 + V_2 - V_\textrm{d}}{2}.
    \label{equ:post_parallelization_voltage_resistive}
\end{aligned}
\end{equation}
When parameters meet $R^2 < \frac{8L}{C}$, i.e., the inductance dominates the interconnection, the paralleling current has a solution of
\begin{equation}
    i(t) = \frac{V_1 - V_2 - V_\textrm{d}}{L \beta}\, e^{\alpha t}\, \text{sin}(\beta t),
\end{equation}
where 
\begin{equation}
\begin{aligned}
    & \alpha = -\frac{R}{2L}, \\
    & \beta = \sqrt{\frac{2}{LC} - \frac{R^2}{4L^2}}.
\end{aligned}
\end{equation}
This model describes a decaying oscillation dynamics, where the first current zero crossing happens at 
\begin{equation}
    t_{\infty} = t|_{i = 0} = \frac{\pi}{\beta}, 
\end{equation}
which is also the time when the parallelization ends.
We can obtain the capacitor voltage after parallelization as
\begin{equation}
\begin{aligned}
    & v_{21}(\infty) = V_1 - (V_1 - V_2 - V_\textrm{d}) \frac{1 + e^{\frac{\alpha \pi}{\beta}}}{2}, \\
    & v_{22}(\infty) = V_2 + (V_1 - V_2 - V_\textrm{d}) \frac{1 + e^{\frac{\alpha \pi}{\beta}}}{2}.
    \label{equ:post_parallelization_voltage_inductive}
\end{aligned}
\end{equation}

\subsubsection{Energy Loss}
The energy loss caused by the parallelization can be calculated from the difference of energy stored in the capacitors per
\begin{equation}
    \Delta E = \left(\frac{1}{2} C V_1^2 + \frac{1}{2} C V_2^2 \right) - \left(\frac{1}{2} C v_1(\infty)^2 + \frac{1}{2} C v_2(\infty)^2 \right)\!,
\end{equation}
where $v_1(\infty), v_2(\infty)$ are the final voltages of the capacitors after paralleling.
For bidirectional parallelization, such as in CH2B circuits, the energy loss is
\begin{equation}
\begin{aligned}
    \Delta E_{\text{CH2B}} &= \frac{1}{2} C V_1^2 + \frac{1}{2} C V_2^2 - C\left(\frac{V_1 + V_2}{2}\right)^2 \\
    &= \frac{1}{4}C(V_1 - V_2)^2.
\end{aligned}
\end{equation}
This energy loss is independent of circuit parameters.

For the DiSeP circuit, the energy loss with resistive interconnection type can be calculated based Equation (\ref{equ:post_parallelization_voltage_resistive}) as
\begin{equation}
\begin{aligned}
    \Delta E_{\text{DiSeP-R}} &= \frac{1}{2} C V_1^2 + \frac{1}{2} C V_2^2 - \frac{1}{2}v_{11}(\infty)^2 - \frac{1}{2}v_{12}(\infty)^2 \\
    &= \frac{1}{4}C\left((V_1 - V_2)^2 - V_\textrm{d}^2 \right)\!, 
\end{aligned}
\end{equation}
which is only affected by the forward voltage drop of the diodes.

The inductive interconnection as in Equation (\ref{equ:post_parallelization_voltage_inductive}) in contrast exhibits an energy loss of 
\begin{equation}
    \begin{aligned}
        \Delta E_{\text{DiSeP-L}} &= \frac{1}{2} C V_1^2 + \frac{1}{2} C V_2^2 - \frac{1}{2}v_{21}(\infty)^2 - \frac{1}{2}v_{22}(\infty)^2 \\
        &= \frac{1}{4} C (V_1 - V_2 - V_\textrm{d})^2 \left(1 - e^{\frac{2\alpha \pi}{\beta}} \right) \\
        &\quad \times \left(1 + \frac{2V_\textrm{d}}{(V_1 - V_2 - V_\textrm{d}) \left(1 - e^{\frac{\alpha \pi}{\beta}} \right)} \right)\!.
    \end{aligned}
\end{equation}
This energy loss can be adjusted through parallel loop circuit configurations.
Table \ref{tab:comparison_paralleling_results} summarizes the above results.

\subsection{Variation of Inter-Module Connection}
We can engineer the inter-module connection to improve circuit performance, as indicated in Figure \ref{fig:disep_structure}. The most common approach is to insert magnetics to control the differential mode inductance \cite{zhang2024analytical}. For bidirectional parallel MMC circuits such as CH2B, this practice brings down the amplitude of the current surge. For the proposed DiSeP topology, the inter-module inductance can further modulate the parallelization dynamics and reduce loss, as described in Table \ref{tab:comparison_paralleling_results}.

\subsection{Modulation Scheme}
\begin{figure}
    \centering
    \includegraphics{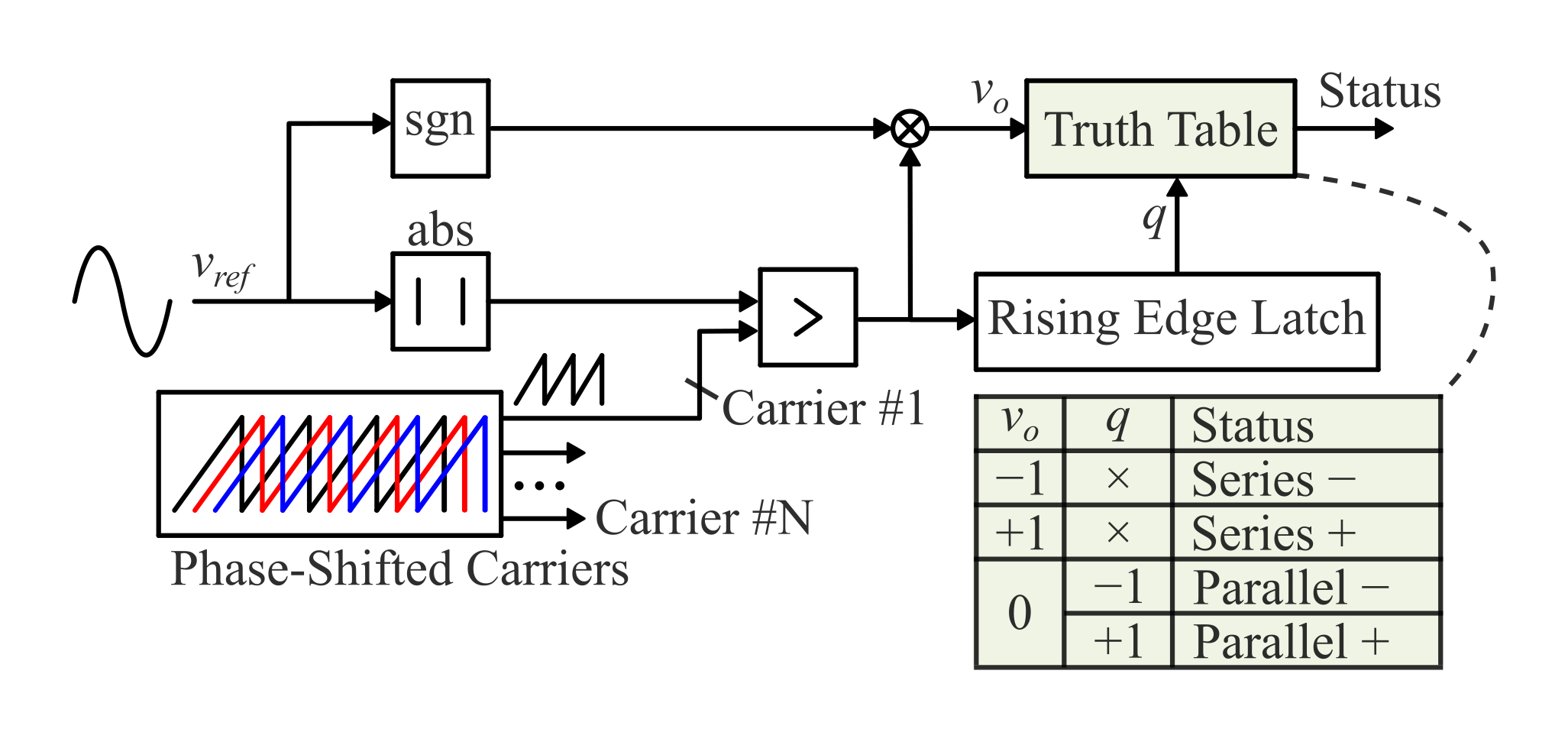}
    \caption{Adaptation of phase-shifted-carrier (PSC) pulse-width modulation (PWM) to DiSeP circuits. In addition to remapping bypass to parallel modes, this modulation scheme implemented a rising edge latch, which takes the absolute value of the modulated voltage as input, to alternate the two parallel modes.}
    \label{fig:control_diagram}
\end{figure}
The DiSeP circuit is compatible with most existing modulation methods, such as phase-shifted-carrier (PSC) and level-shifted-carrier (LSC) pulse-width modulation (PWM). We adapt the PSC-PWM scheme to the DiSeP circuit in Figure \ref{fig:control_diagram}. Similar to CH2B circuits, the modulator preferably switches to one of the parallel modes when the quantizer generates a zero, instead of a bypass mode as in series-only MMC circuits. The only extra effort for the DiSeP circuit is to alternate the two parallel modes to guarantee a bidirectional parallelization, which is achieved through a simple rising-edge latch as shown in Figure \ref{fig:control_diagram}. This latch takes the absolute value of the modulated voltage as input, which in principle serves as a separator between \emph{Series} modes and others. This approach guarantees that the polarity of parallel modes is flipped every time, and thus alternates between \emph{Parallel--} and \emph{Parallel+}.

\section{Results}
We initially discuss several general design considerations, particularly the paralleling situations with different circuit configuration. We subsequently introduce an experimental prototype and its test platform. Followed is the analysis of the circuit performance.

\subsection{Pre-Prototyping Analysis}
\begin{figure}[t]
    \centering
    \includegraphics[trim=0 4mm 0 0,clip]{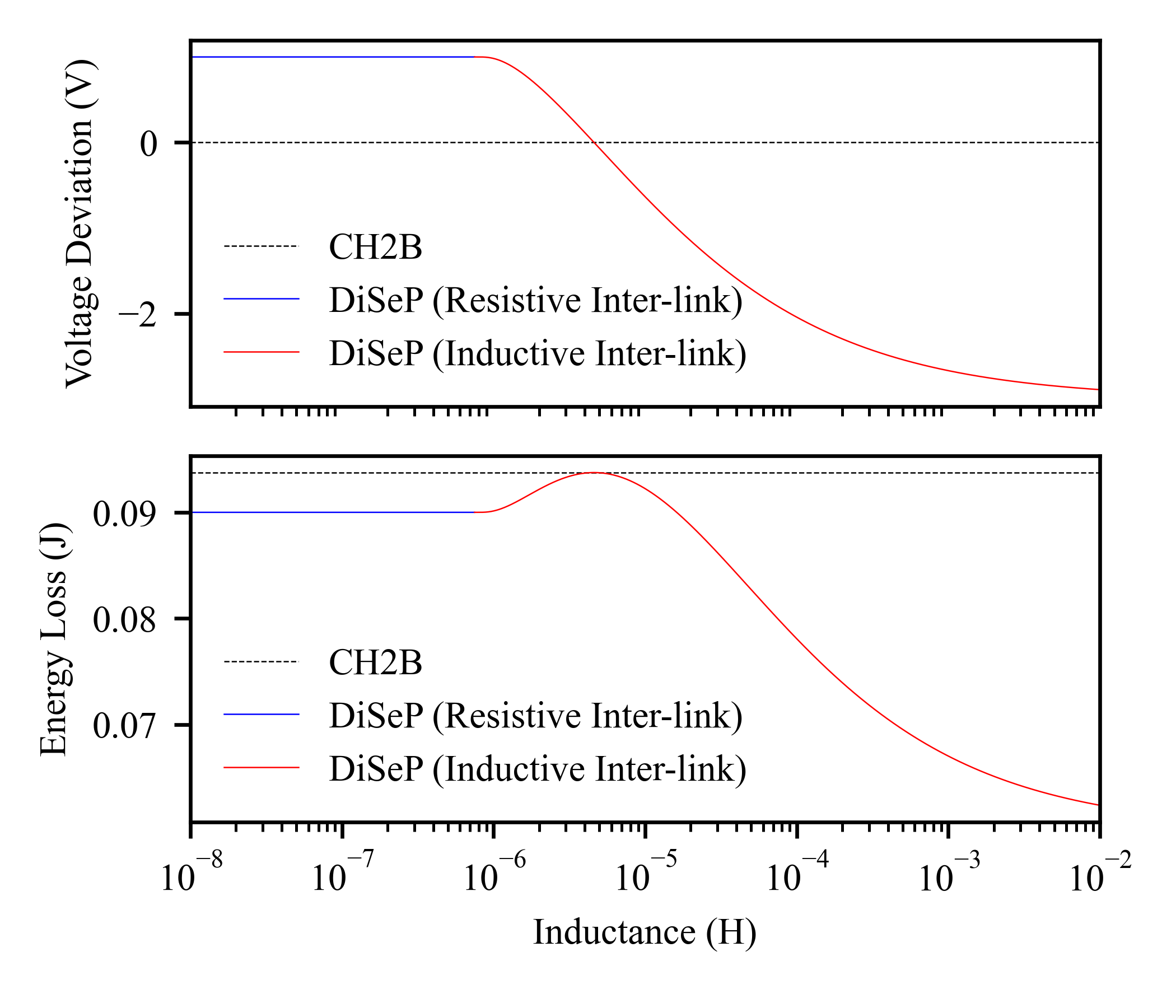}
    \caption{Module balancing for a varying differential mode inductance. Starting with resistive behavior, where the parallelization results in a constant voltage deviation and energy loss, increasing inductance pushes the parallelization into the inductive area, where the voltage deviation decreases. A CH2B circuit with the same circuit configuration is provided for comparison.}
    \label{fig:paralleling_results}
\end{figure}
\begin{figure}
    \centering
    \includegraphics[trim=0 4mm 0 0,clip]{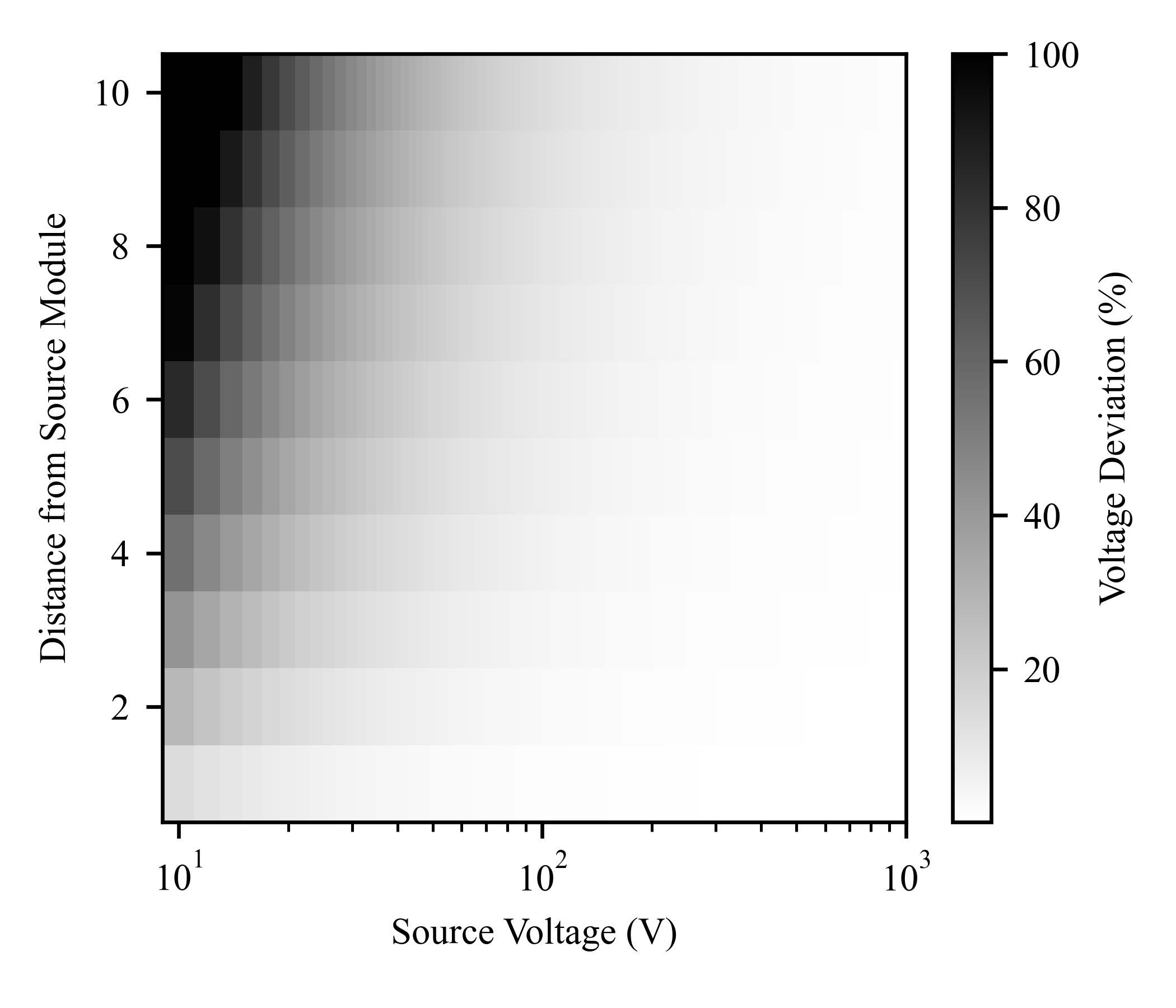}
    \caption{Voltage deviation as a function of distance from the voltage source and its significance under different application voltage.}
    \label{fig:voltage_deviation_colormap}
\end{figure}
With a parallel loop model of 15,000\,{}µF capacitor per module and 20\,{}m$\Omega$ resistance as well as an assumption of a 10\,{}V voltage difference, we explore the influence of differential-mode inductance on parallelization. Varying from 10 nH to 10\,{}mH, the parallelization dynamics transition from resistive to inductive (Figure \ref{fig:paralleling_results}). When behaving resistive, the parallelization results in a constant voltage difference of one diode forward voltage drop. Entering the inductive region, beyond 0.8\,{}µH in this case, the voltage difference after parallelization varies accordingly. The voltage difference decreases as the inductance increases, and can further become negative, where the module of lower voltage before paralleling ends up with a higher voltage after. Moreover, we can achieve a zero voltage difference by controlling the inductance, i.e., 4.5\,{}µH in this example.

Figure \ref{fig:paralleling_results} also compares the parallelization energy loss. In addition, in the case where the inductance is adjusted to achieve zero voltage difference, the energy loss of DiSeP is consistently lower than for a CH2B circuit. The resistive inter-link configuration constantly reduces the paralleling loss by 4\%, whereas a proper inductance can further reduce the loss.

In this paper, we want to challenge the circuit in the hardest configuration for the parallelization function, which feeds all power through a local dc link into one module only \cite{zhang2023gallium}. Thus, the entire power has to be distributed through parallelization to all modules. This operation mode challenges the parallel mode and furthermore the diode path and associated voltage drop (Figure \ref{fig:voltage_deviation_colormap}). 
Thus, the post-parallelization voltage of modules depends on the topological distance (number of modules through which the supply power has to pass) from the module with the dc supply and decreases linearly. These drops are on the order of diode forward voltages, which are typically small relative to typical module voltages in most grid-level MMCs and likewise present in conventional MMCs with IGBTs.
Proper differential-mode inductance can further reduce or eliminate voltage differences if they exceed the tolerable  range (Figure \ref{fig:paralleling_results}). The inter-module differential-mode inductance can also serve for over-compensating, i.e., boosting the voltage.

\subsection{Experimental Prototype and Platform}
\begin{figure}
    \centering
    \includegraphics{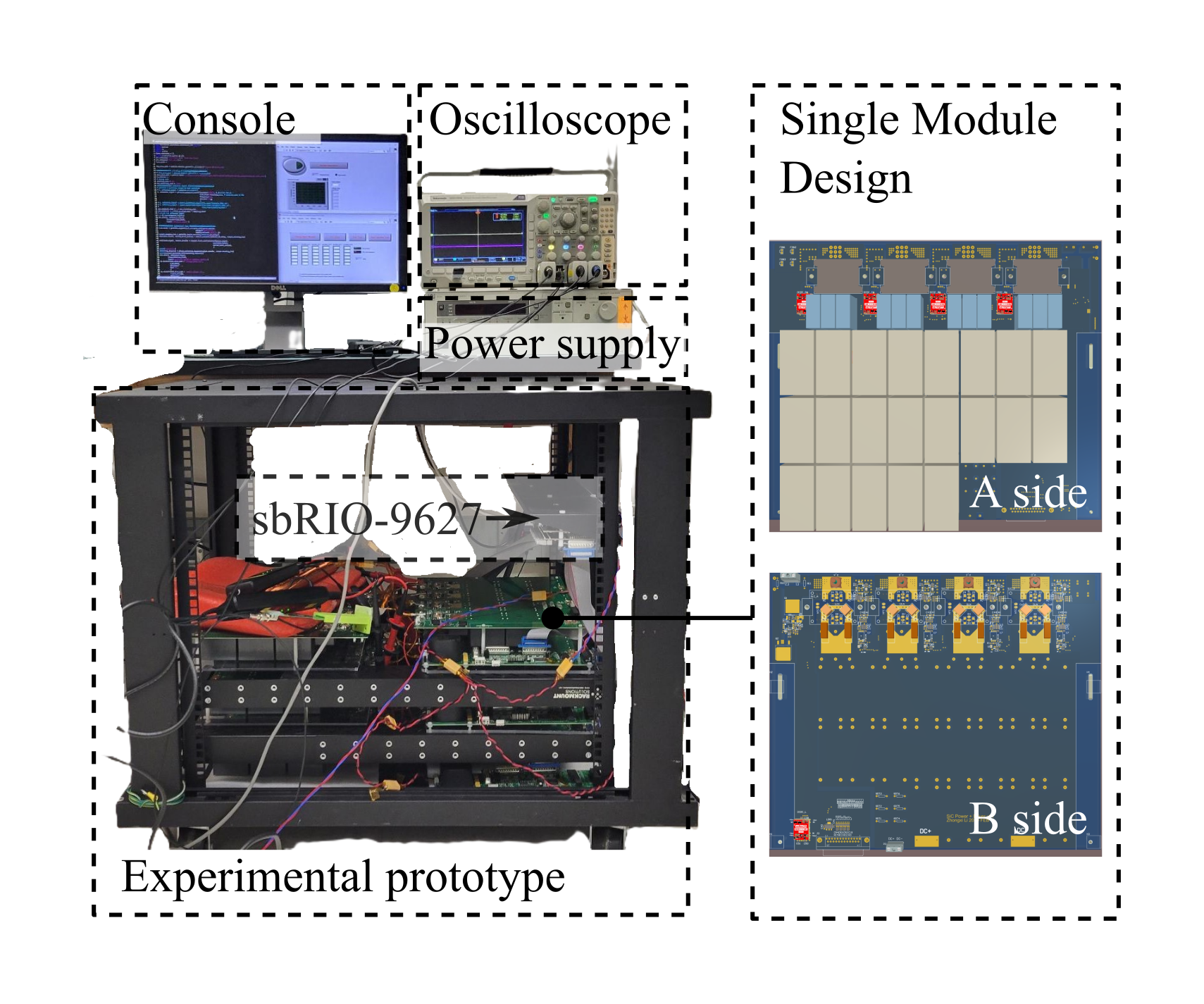}
    \caption{The six-module DiSeP experimental prototype and the test platform. The carrier frequency for modules is set to 10\,kHz. A programmable dc voltage source feeds the capacitor of the third module.}
    \label{fig:photo_prototype}
\end{figure}
We built a six-module experimental prototype with 8 m$\Omega$ on-resistance FETs, 1.2 V forward-voltage diodes, and a capacitor bank of 15,000\,{}µF (Figure \ref{fig:photo_prototype}). We did not insert dedicated magnetics between modules beyond parasitics. A programmable dc voltage source with a nominal voltage of 35\,{}V supplies the third module, which has to distribute the power to the whole system. An embedded controller (sbRIO-9627, National Instruments) operates the prototype with 40\,{}MHz bandwidth following the algorithm of Figure \ref{fig:control_diagram}. The individual switching rate, i.e., carrier frequency of PSC-PWM, is set to 10\,{}kHz. A 200\,$\Omega$ resistor loads the prototype, which generates a sinusoidal output with an amplitude of 200\,{}V and frequency of 60\,{}Hz.

The experiment was measured with  voltage probes (100\,{}MHz bandwidth, THDP0100, Tektronix) and a Rogowski current probe (20\,{}MHz bandwidth, CWTMini, PEM). All data were collected through a digital oscilloscope (2.5 GS/s sampling rate, MDO3054, Tektronix).

\subsection{Circuit Performance}
\begin{figure}
    \centering
    \includegraphics{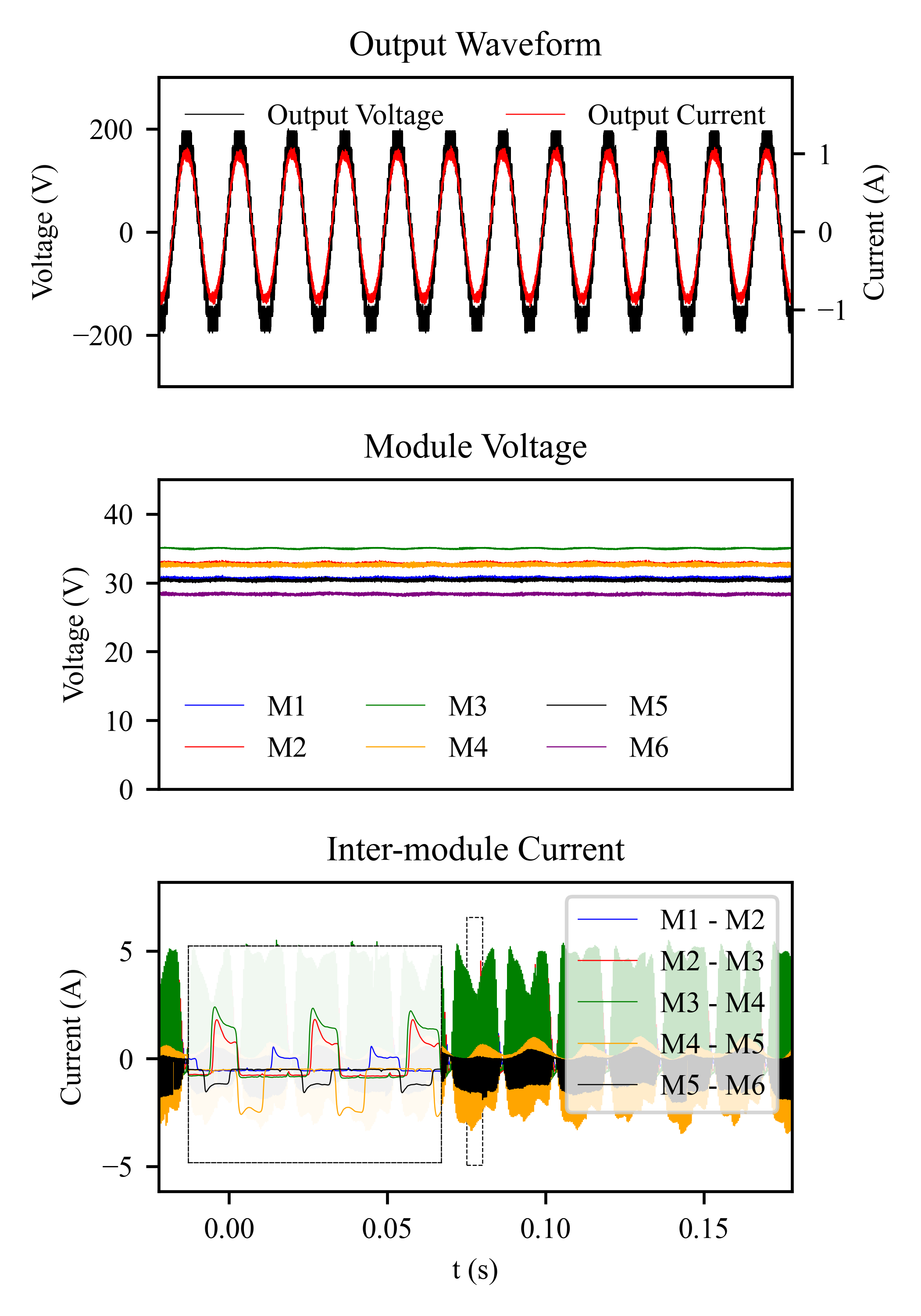}
    \caption{Measured output voltage and current, the voltages of individual modules, and the current of the inter-module connections. The output voltage has an amplitude of 200\,{}V at a current of 1.0\,{}A. The first to the sixth modules maintain their voltage at respectively 30.7\,{}V, 32.7\,{}V, 35.0\,{}V, 32.6\,{}V, 30.4\,{}V, and 28.4\,{}V, which reflects the supply in Module 3 and the diode voltage drops. The inter-connection current peaks at 5.1\,{}A with a unipolar decaying shape.}
    \label{fig:output_waveforms}
\end{figure}
\begin{figure}[t]
    \centering
    \includegraphics{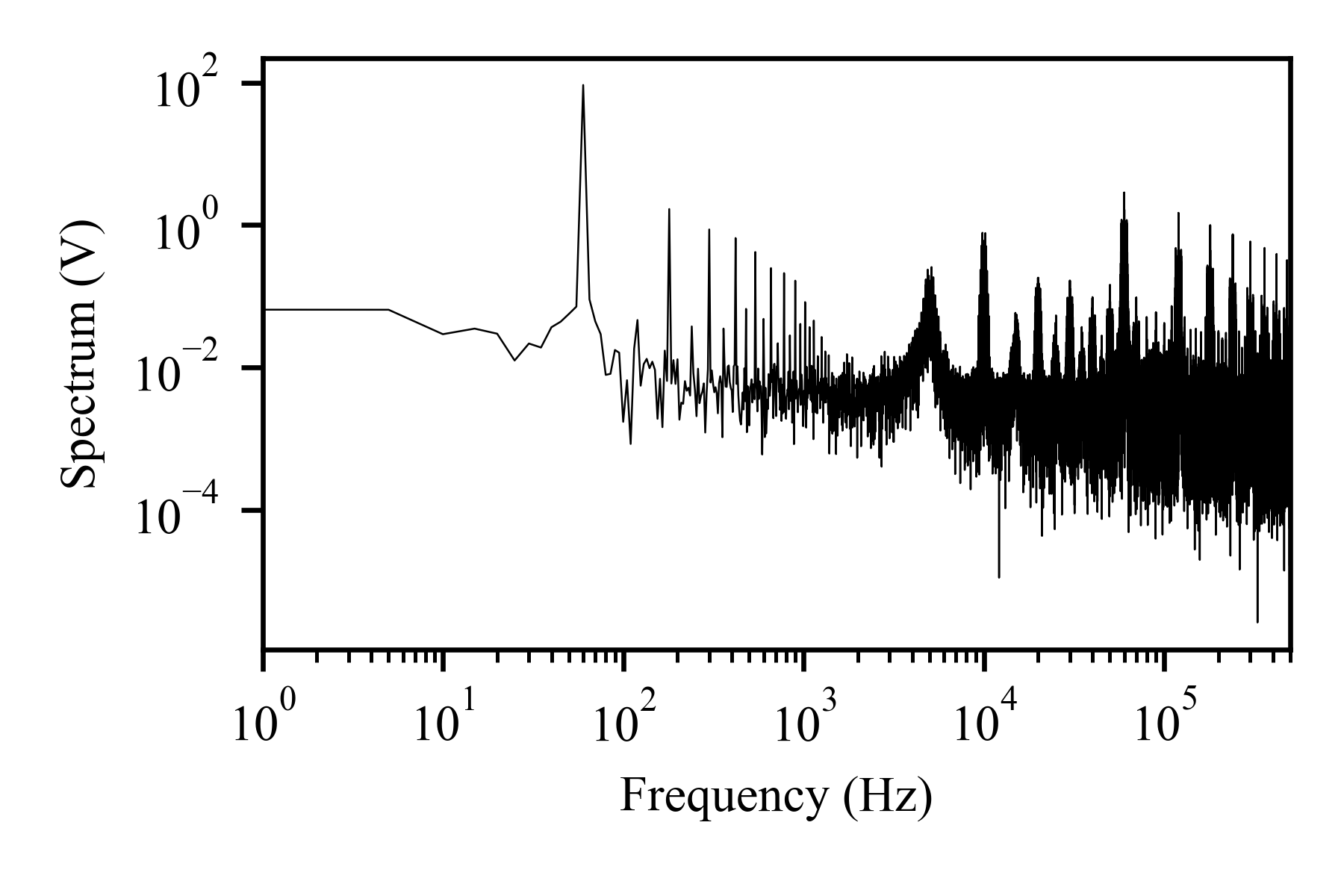}
    \caption{Normalized spectrum  of the output voltage (voltage THD of 9.4\% and a THD+N of 10.3\%).}
    \label{fig:output_spectrum}
\end{figure}
\begin{figure}
    \centering
    \includegraphics{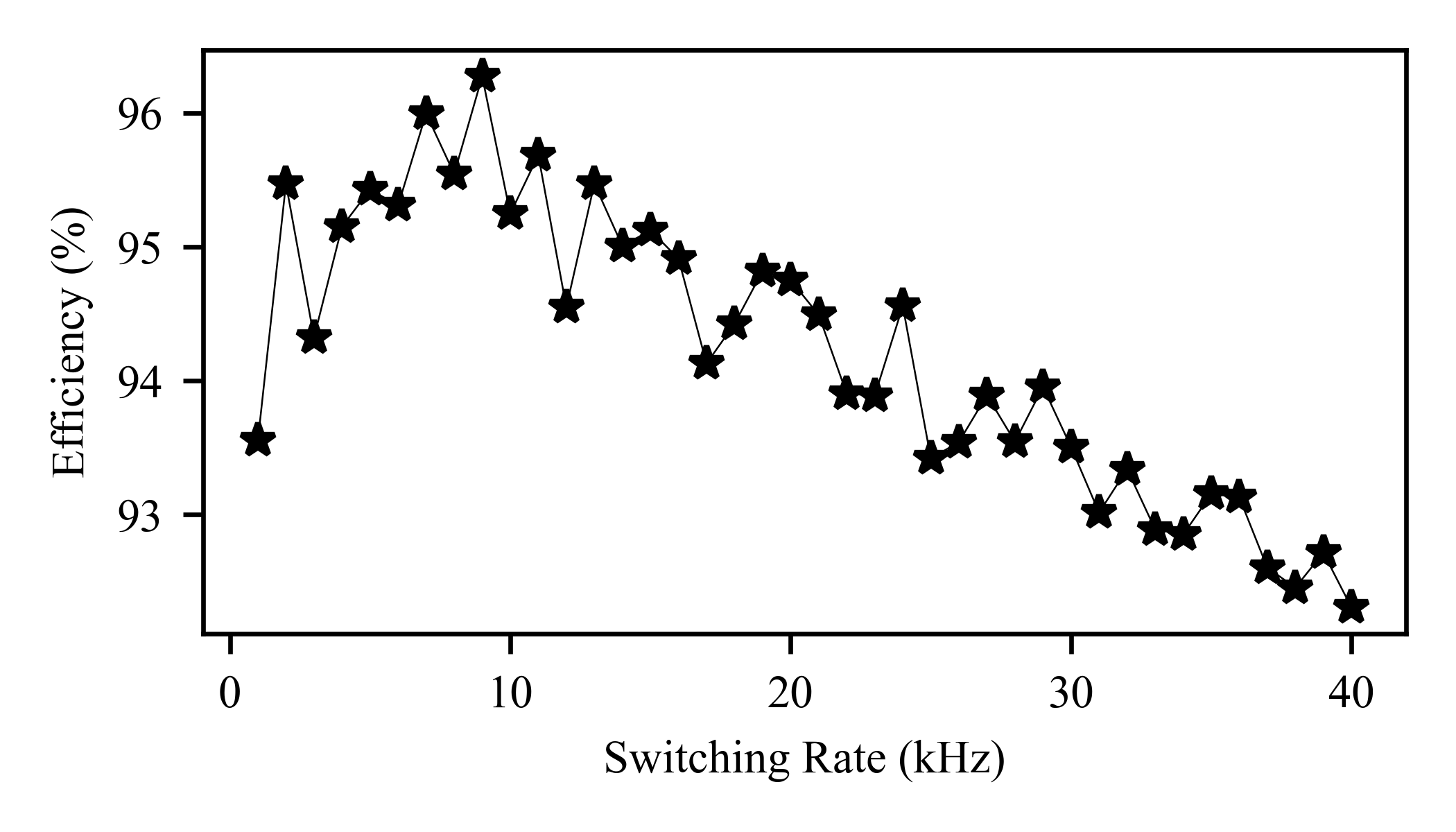}
    \caption{Efficiency variation for different switching rates. starting with 93.6\,\% at 1\,{}kHz, the efficiency increases with the switching rate, reaches a peak of 96.3\,\% at 8\,{}kHz, and drops with further increase of the switching rate.}
    \label{fig:efficiency_curves_fsw}
\end{figure}

We summarized key waveforms of the circuit in Figure \ref{fig:output_waveforms}, particularly the output voltage and current, individual module voltage, and the inter-module current. The output voltage has an amplitude of 200\,{}V, which leads to a current amplitude of 1.0\,{}A and an output power of 100\,{}W.

Figure \ref{fig:output_spectrum} provides the spectrum of output voltage, calculated by fast Fourier transform. The distortion occurs mainly at the harmonics of the output frequency (i.e., 120\,{}Hz, 240\,{}Hz, etc.) and of the switching frequency (10\,{}kHz, 20\,{}kHz, 40\,{}kHz, etc.). The output voltage exhibits a total harmonic distortion (THD) of 9.4\,\%, and a total distortion (TD = THD+N) of 10.3\,\%.

Figure \ref{fig:output_waveforms} also presents waveforms of the individual module voltage and inter-module current. Compared with the 35\,{}V voltage source connected to the third module, the second and fourth modules stabilized their voltage at two diodes' forward voltage drops below the input, respectively 32.7\,{}V and 32.6\,{}V. The first and fifth modules show a voltage deviation of four diode drops o that they respectively reach 30.7\,{}V and 30.4\,{}V. The sixth module, the furthest one from the power supply, maintains its voltage at 28.4 V. These observations align accurately with our analysis in Figure \ref{fig:paralleling_results} and Table \ref{tab:comparison_paralleling_results}.

The inter-module current peaks at 5.1\,{}A. The zoomed-in waveform, which shows a decaying current shape, validates the assumption of a resistive interconnection. Further, with properly engineered differential-mode inductances, we can reduce the current surge as well as reduce or eliminate voltage deviation. 

We further explored the influence of switching rates on the circuit efficiency (Figure \ref{fig:efficiency_curves_fsw}). Previous studies found that a higher switching rate leads to a higher switching loss but can reduce the parallelization loss \cite{6763109}. The performance of this prototype aligns with the previous findings from CH2B. At a switching rate of 1\,{}kHz, the converter reaches an efficiency of 93.6\,\%. Increasing the switching rate significantly reduces the paralleling loss and raises the efficiency, up to a peak value of 96.3\,\% at 8\,{}kHz. At the nominal switching rate of 10\,{}kHz, the efficiency is 95.7\,\%. Further increase of the switching rate drives the switching loss beyond the savings of the paralleling loss so that the efficiency decreases (92.3\,\% at 40\,{}kHz).

\section{Conclusion}
This paper proposes a direction-selective parallel (DiSeP) topology for modular multilevel and cascaded-bridge converter applications. This topology can generate bipolar series modular output and achieve sensorless voltage balancing as well as reduced impedance. We illustrated its working principle and analyzed its key features, including impedance and parallelization. We validated our theories with a high-quality high-efficiency experimental prototype.

This module type is the first-ever four-transistor solution to achieve sensorless voltage balancing and bipolar output. The DiSeP circuit can serve as an alternative to the H-bridge module and achieve sensorless module balancing with a marginal extra cost on diodes. DiSeP is also a strong competitor to the CH2B circuit as it significantly cuts the cost on transistors and their auxiliary circuits while keeping key functionalities. In summary, the proposed DiSeP topology holds great potential in a variety of energy applications due to its powerful capabilities, high cost-efficiency and high software reliability.

\bibliographystyle{Bibliography/IEEEtranTIE}
\bibliography{Bibliography/IEEEabrv,Bibliography/main}\ 

\vspace{-1cm}
\begin{IEEEbiography}[{\includegraphics[width=1in,height=1.25in,clip,keepaspectratio]{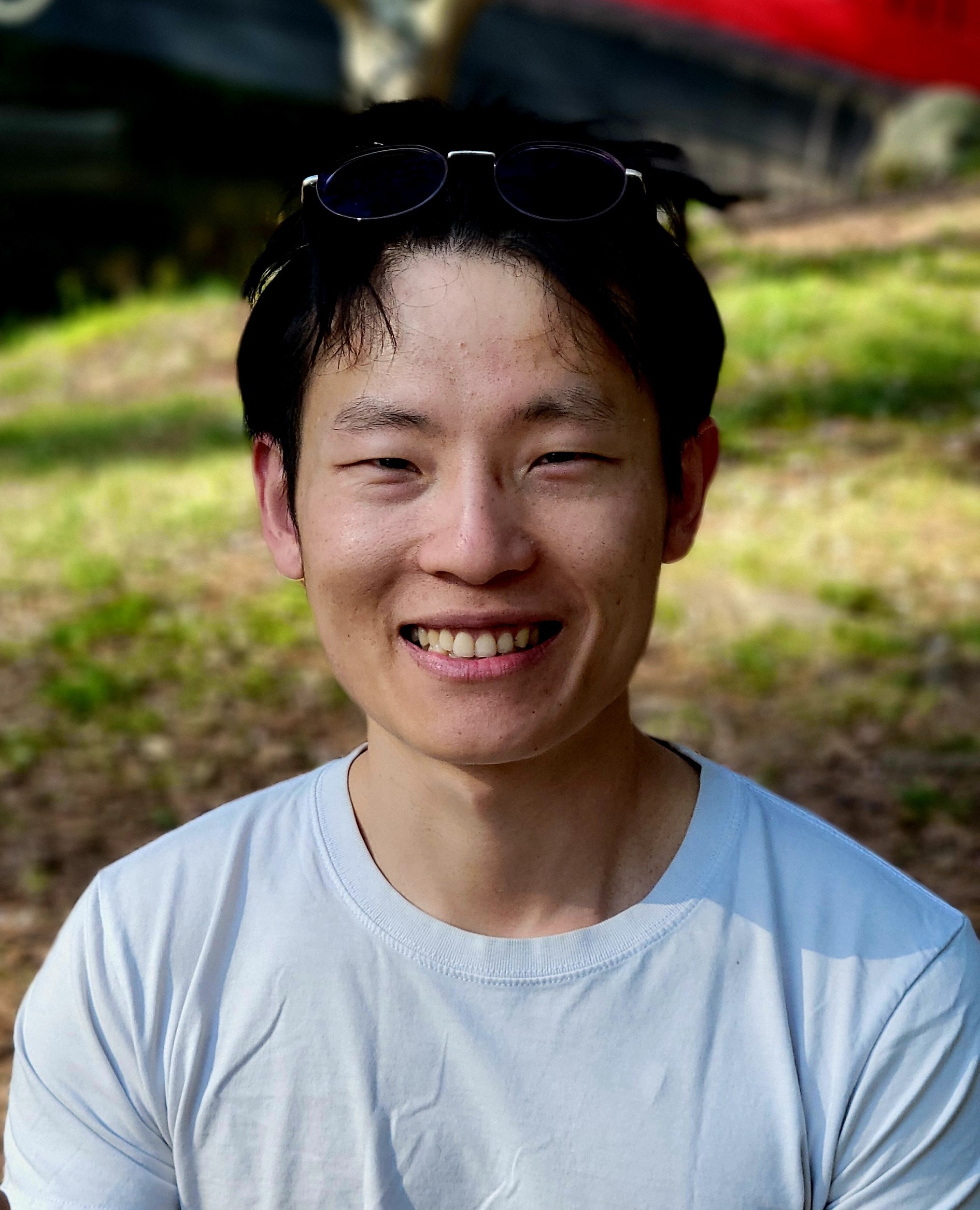}}]
{Jinshui Zhang} earned his bachelor's degree in electrical engineering from Tianjin University, Tianjin, China, in 2018 and his master's degree in electrical engineering from Xi'an Jiaotong University, Xi'an, China, in 2021. Currently, he is pursuing a doctoral degree of Electrical and Computer Engineering at Duke University, Durham, NC, USA. 
\end{IEEEbiography}

\vspace{-1cm}
\begin{IEEEbiography}[{\includegraphics[width=1in,height=1.25in,clip,keepaspectratio]{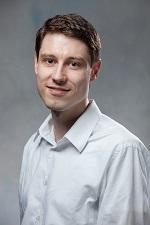}}]
{Stephan M. Goetz (Member, 2014)} received his undergraduate and graduate degrees from Technische Universit at Munchen (TU Muenchen), Munich, Germany, and Ph.D. training from TU Muenchen and Columbia University, New York, NY, USA, with a thesis on medical applications of power electronics. He previously worked in the automotive industry in various positions and levels with a focus on electric drive trains, machines, power electronics, and vehicle architecture. He, furthermore, developed automotive chargers and charging technologies, which are installed on vehicles and in charging stations around the world. He has worked on grid integration and grid-stabilizing functions of chargers. His research interests include high-quality, high-power, high-bandwidth electronics and magnetics for drive and medical applications, as well as integrative power electronics solutions for microgrids and electric vehicles.
\end{IEEEbiography}

\end{document}